# Improved simulation method for the calculation of the intrinsic viscosity of some dendrimer molecules

*Esteban Rodríguez, Juan J. Freire*[*]

Departamento de Ciencias y Técnicas Fisicoquímicas, Facultad de Ciencias,
Universidad Nacional de Educación a Distancia (UNED), Senda del Rey 9, 28040
Madrid, Spain

and

*G. del Río Echenique,  J.G. Hernández Cifre,  J. García de la Torre*

Departamento de Química Física, Universidad de Murcia,
30071 Murcia, Spain

[*]Corresponding author. Tel/fax: +34-913-944-154
Electronic mail: jfreire@invi.uned.es

Short Title: Simulation of the Intrinsic Viscosity of Dendrimers

**Abstract**:

A method previously proposed for calculating the radius of gyration and the intrinsic viscosity of dendrimers is modified to give a more accurate description of existing experimental data. The new method includes some features that were not previously considered, namely: a) a correction term to take into account the contribution of individual friction beads, whose volumes are not negligible in comparison with the molecule size,  b) a realistic distribution of internal angles between successive beads that define branching points in the molecule, c) a distribution of distances between branching points computed from Molecular Dynamics simulations of a small dendrimer with explicit solvent. Modification a) alone is able to give a good description of the experimental results obtained for polypropylene-imide with a diaminobutane core in water, while the simultaneous use of the three modifications is needed to adequately describe the experimental data of monodendrons and tridendrons of polybenzylether in THF.

**Introduction**

The peculiar physical properties of dendrimers constitute one of the reasons explaining the current interest for this type of molecules [1-2]. Different theoretical models and simulation methods have been devoted to understand these properties. Thus, Monte Carlo [3,4] (MC), Brownian Dynamics [5] and Molecular Dynamics (MD) methods [6-8] have been applied to different ideal and atomistic representations of the dendrimer molecules. From the experimental point of view, a particularly dramatic effect is observed in the experimental data of the intrinsic viscosity, $[\eta]$. This property is usually employed as a standard characterization technique in conventional polymers due to its simplicity from the experimental point of view and its direct relation with the molecule size. In the case of dendrimers, however, the variation of $[\eta]$ with molecular weight, or with the generation number, *g*, often exhibits a maximum, whose location depends on the dendrimer chemical structure. In a naïve description, the maximum is correlated with a minimum in density after which the dendrimer cannot grow without a significant congestion of units. However, a precise location and quantitative prediction of the viscosity maximum also has to take into account the more subtle variation of the solvent permeability in these relatively small molecules. Therefore, a simple lattice or bead chain model, similar to those used to describe universal properties in polymer systems, is not able to give an acceptable quantitative description of the experimental viscosities for the available range of values of *g* unless some information about structural details of the real molecules is incorporated.

In previous work [9], we proposed a computational method based in a model for Monte Carlo (MC) simulations that considers a bead for each branching point and an additional bead in the middle point between them. Since the main aim of the present publication is to incorporate some improvements to this method, we describe it briefly now. The beads positions are changed by a "single-bead jump" algorithm in which the distances between neighboring units and neighboring branching units follow a broad distribution in a range of realistic values. The limits of this range are set according to the results obtained by relatively short Molecular Dynamics (MD) simulations previously performed for an atomistic model. (It is shown that these MD simulations would require a much longer computer time to directly provide accurate viscosity results). The model also introduces a rigid-spheres potential to avoid overlapping of non-neighboring beads in terms of a distance σ, which is set to yield the closest reproduction with the MC samples of the existing data for the mean radius of gyration, $Rg$, for a given set of dendrimer molecules and different generation numbers [1,10,11].

The intrinsic viscosity calculations are performed using the "lower bound" computational scheme devised by Fixman [12]. In these calculations, we assume that the molecules are constituted by friction beads centered in the branching point units. A value for the friction radius of the bead, $f_r$, is needed to describe hydrodynamic interactions. This parameter is set to give the closest quantitative reproduction of the experimental viscosity data (though this fit alone is unable to warrant a good reproduction on the maximum location for this property and sometimes the quantitative agreement can be only limited to a narrow range of generation numbers). In the lower bound calculations, the results are obtained by averaging some intermediate magnitudes over a statistical sample. The sample is constituted by randomly selected conformations that were obtained in the previous MC simulations described above. Alternatively, it is possible to use a Kirkwood-Riseman calculation of the intrinsic viscosity for instantaneously rigid conformations. The results would provide an upper-bound of this property when averaged on a similar conformational sample. Differences between the upper and lower bound results are usually small, but the lower bound method is significantly more efficient from the computational point of view for the treatment of many-bead flexible molecules [2].

Our multi-step method was able to give a fair quantitative description of the intrinsic viscosity data of some dendrimer molecules. Thus, for polyamidoamine dendrimers with a ethylendiamine core (PAMAM-EDA), the simulation results were compatible with the somehow scattered experimental data of these dendrimers in water [1] (the relatively large non-systematic changes in the variation of the $[\eta]$ vs $g$ data reported in the bibliography do not allow us to expect that a better agreement can be reached with any improved model. Consequently, these dendrimers are not considered in the present work). For the case of polypropylene-imide with a diaminobutane core (PPI-DAB) in water we considered a set of data [10] that includes four generation numbers, not far from the results obtained independently for the same dendrimers in methanol [13]. Also, two additional bibliographic data corresponding to successive generations have been reported [14]. The latter data are in disagreement with the larger set and show a difference between the values higher than in any other available set of $[\eta]$ data for dendrimer systems. The simulation results showed a qualitative description of the more consistent experimental set of data, but with significantly smaller quantitative values. Consequently, we were nor able to offer an adequate agreement with the experimental data for the PPI-DAB case.

The best performance of our previous method was achieved for the case of monodendrons and tridendrons of polybenzylether (PBzE). We performed a comparison with the consistent experimental set of $[\eta]$ vs $g$ data obtained time ago for tetrahydrofuran (THF) solutions [15] and a reasonable agreement was obtained for both types of molecules. In spite of this general agreement, some consistent differences between the experimental and simulation data still remained. Thus, the simulation results clearly overestimated the experimental data for the smallest values of $g$. Moreover, the same overestimation is also shown for the highest $g$ in the case of the tridendrons, where the viscosity peak is more prominent. These differences indicate the convenience of devising an improved method, where more details of the molecular structure are incorporated.

In this work, we describe some refinements with which we have been able to achieve a significantly better agreement between experimental and simulation data for PPI-DAB in water and for the mono and tridendrons of PBzE in THF. In a first stage, we have introduced a simple correction to the simulation data, based in adding a term representing the contribution of an individual friction bead. With this correction, the intrinsic viscosity of a single bead is not zero but has the value corresponding to a rigid sphere with the chosen friction radius. Details on the justification of this correction as the $0^{th}$-order term to the conventional Kirkwood-Riseman calculation of the intrinsic viscosity, which accounts for hydrodynamic interactions with terms of order $-1$ in interbead distances, can be found elsewhere [16]. Some kind of correction of this type should actually be introduced in all systems but the correction is commonly ignored in the case of long polymer chains, where the hydrodynamic volume is much higher than the beads individual contribution. The correction should obviously be also pertinent for our results which provide a lower-bound estimation for the considerable smaller dendrimer molecules but it was ignored in our previous treatment. We anticipate that this modification alone is adequate to correctly describe the PPI-DAB data.

In our previous simulations we obtained a fair agreement between experimental data and our previous MC results for the PBzE dendrimers. However, we will see that the introduction of the single bead correction actually worsens this agreement, specially in the case of tridendrons. Therefore, assuming that a bead correction is pertinent for all types of small dendrimer structures, it is apparent that a good description of the PBzE structures should incorporate further details to the molecular model. These molecules have particularly rigid connections between their branching points. Consequently, as a first refinement in the MC model should include a realistic distribution of the internal angle $\Theta$, formed by the connections between successive branching points. This improvement will be proved to be sufficient to provide again a fair agreement between MC and experimental results. Furthermore, we have also considered that the large phenyl groups are conditioning the conformational characteristics, particularly the distribution of distances. Since the interactions between phenyl groups can noticeable change in different solvents, we have obtained a new distribution of distances between neighboring branching units (or branching units and ends) from MD simulations that explicitly include solvent THF. (Previously, we obtained the distance distribution from MD simulations performed for single molecules without solvent.) With all these new features incorporated to the model, we have finally been able to obtain a good description of the radius of the experimental data of PBzE dendrimers.

**Computational methods**

Our original method was previously detailed [9] and was summarized above. It incorporates data to describe the range of realistic distances between neighboring units and also between neighboring branching units. These data were obtained from MD simulations for single dendrimer molecules. A description of the particular chemical structures of the considered molecules, and details on the choice of units and the (non-standard) definition of the generation numbers that are also employed in the present work can also be found in [9].

For models with identical beads as in our case, the simplest correction for the contribution of individual beads to the intrinsic viscosity [16] consists in adding the viscosity of a single rigid sphere obtained from the Einstein formula to the "raw" simulation viscosity, $[\eta]_{sim}$, which we computed with the lower bound method. This way

$$[\eta]_{corr} = [\eta]_{sim} + (10/3)\pi f_r^3 / m_u \qquad (1)$$

where $m_u$ is the mass of the unit, obtained from the molecular weight of the chemical groups engulfed by each friction bead. $m_u$ can be directly computed from total number of friction units ($N$-$N_i$) and the molecular mass $M$ of the dendrimers reported in Table 1 of [9]. Therefore, the procedure to calculate the $[\eta]_{corr}$ values is: 1) to select a value of the friction radius, 2) to compute $[\eta]_{sim}$ with the lower bound approximation, i.e. performing the adequate averages from a statistical sample extracted from the Monte Carlo simulation, and 3) to add the individual bead correction, contained in Eq. (1), which is slightly changing with the generation number for a given type of dendrimer. This process has to be repeated with several $f_r$ values until the best description of the experimental data (at least in a particular range of values of $g$) is achieved. Therefore, the MC configurational samples obtained in our previous work can be employed again to obtain $[\eta]_{sim}$ and $[\eta]_{corr}$ and new simulations are not required if this correction alone is able to reproduce the experimental data. In theory, a similar correction term for the individual bead should be added to the simulation data to calculate the corrected squared radius of gyration [17]. However, we have verified that this correction is significantly smaller than in the intrinsic viscosity case and, actually, it is practically irrelevant for the choice of distance $\sigma$ that best reproduces the existing experimental data of $R_g$.

In our previous calculations [9], the MD simulations covered several generation numbers and were performed for single molecules. Since flexible molecules in the vacuum tend to collapse, we had to impose a reduction of the potential cut-off to avoid this trend [18]. In order to incorporate solvent effects to our model, we have performed new MD simulations for the system constituted by the small monodendron of PBzE with $g$=2 immersed in a bath of THF molecules. With this aim, we use the "Amorphous Cell" module of the "Materials Studio" software package [19]. With this software, we prepare a simulation box with the selected $g$=2 dendrimer and 50 solvent molecules. The box size is fixed to yield a system density of 0.9 g./c.c. similar that of THF at room temperature. The resulting system is conveniently thermalized with a series of short MD runs at different temperatures. Finally we perform the useful MD run over $4 \times 10^6$ steps, equivalent to 4 ns, to analyse properties. The MD runs are performed with module "Discover". For the MD simulations we have used the same procedures and parameters described in our previous work [9], except in the case of the potential cut-off value. Now, we employ the relative high fixed cut-off default value of 9.5 Å, instead of the lower fitted values previously used for the single molecules. We have computed the distance distributions over 400 frames (each one collected after $10^4$ MD steps). The distance distribution functions for the g=2 monodendrons have been obtained averaging over the 4 distances joining the non-central branching units and the end units.

In addition to considering a distribution of distances, the modified model for the PBzE molecules proposed in this work also uses distribution of angles between neighboring branching units, $\Theta$. (In these molecules, the angle formed by two branching units and the middle point between them is rigid). The distribution is estimated from the analysis of MD simulations performed with an atomistic model for the $g$=2-4 dendrimers and it has been computed taking into account the different angles formed by the connected branching points or end units. The estimated results are introduced as input data in a new MC code that takes care of the distribution of angles between neighboring branching points (9 new probability checks per bead move in the most general case of an inner branching unit not located in the dendrimer core).

**Results and Discussion**

The results for the distribution of distances for the $g$=2 monodendron PBzE dendrimer molecule are given in Fig. 1. Consistently with the results previously obtained for larger dendrimers [9], the distance distribution obtained with the single molecule shows a sharp peak close to the upper limit of the distance range. However, the distribution obtained for the system with explicit solvent is more symmetric and flatter.

In Fig.2 we present the distribution of cos($\Theta$) in six equally spaced intervals (equivalent to the mean probability in these intervals). The angle distribution shows a clear peak close to $\Theta=4\pi/3$. It is observed that the results obtained for the angle distribution in single dendrimers with different generation numbers are very close. Actually, we have verified that introducing moderate changes to the angle distributions does not significantly affect the final results. (Only considerably flat distributions yield substantially different Monte Carlo averages.) Therefore, we assume that the variation of the distribution with the generation number, topology, the position of the branching unit inside the molecules or specific details in the MD procedure cannot change much the final results. Consequently, a single angle distribution function is used in all our MC calculations for the PBzE dendrimers. It should be considered that the consideration of different types of angle distributions would add a non-desirable complexity to the model.

As we have explained above, the method that we have previously proposed was able to give a qualitative variation of the intrinsic viscosity vs $g$ similar to that exhibited by the most consistent experimental set of data for PPI-DAB [10]. However, the friction radius employed to obtain this qualitative agreement rendered quantitative simulation values that were consistently and significantly smaller than the experimental values of $[\eta]$ (see open squares in Fig.3). On the other hand, higher values of the friction radius give results quantitatively closer to the experimental data but showing a substantially different trend (open circles in Fig. 3). (This case clearly shows that a simple fit of $r_f$ cannot describe the experimental data unless a correct theoretical procedure is employed.) Accordingly to these conclusions, it seems that the introduction of the correction for individual beads according to Eq. 1 may be particularly useful in this case. In Fig. 3 we also show the results obtained from the same MC samples obtained in [9] for these molecules ($\sigma=4.0$ Å) and application of Eq. (1) with the the value $r_f=2.65$ Å (filled diamonds). A good accordance is observed between the simulation results and this experimental set, in spite that we do not consider angle restrictions and we are not introducing a solvent-dependent distance distribution. Therefore, we conclude that the model employed in our previous work can accurately reproduce the $[\eta]$ vs $g$ behaviour for PPI-DAB and, probably, for any other dendrimer with a not very rigid chain and not too long segment between the branching points, once the individual bead correction term is introduced.

A comparison between the experimental data in THF [15] and different MC results is shown in Fig. 4 for the intrinsic viscosity of the PBzE monodendrons. A similar comparison is included in Fig. 5 for the PBzE tridendrons. The simulation results previously obtained with the original method, which did not included individual bead corrections (open circles in both Figs.), showed a fair agreement with the experimental data using a single value of $r_f$ for monodendrons and tridendrons, $r_f$=4.65 Å. (However, this value is unrealistically high when compared with the range of distances between branching points and some discrepancies with the experiments are observed both for the lowest and highest generation numbers). In Figs. 4 and 5 we also show the results obtained now with the MC samples that we have previously generated ($\sigma$=3.5 Å) and the correction term corresponding to the common friction radius $r_f$=2.65 Å, for monodendrons and tridendrons (open squares). Although the simulation results for monodendrons in Fig. 4 are still in reasonable accordance with the experimental data, it is observed in Fig. 5 that there is a strong disagreement between the general trend of the simulation data and the experiments for tridendrons of high generation number. (This illustrates again that the fit of $r_f$ may not lead to a reasonable agreement between MC results and experimental data unless the correct features are introduced in the numerical method). Therefore, if we want to consistently introduce the individual bead corrections which have been revealed successful for the PPI-DAB case, the molecular model for the PBzE molecules clearly needs of further improvements.

The most significant feature of the angle distributions for the PBzE monodendrons as shown in Fig. 2 is the sharp peak close to $4\pi/3$. Considering the three angles corresponding to a single intermediate branching unit (i.e. a unit not placed in the molecule core or in end groups) the distribution implies that this unit, the preceding unit and the two successive branching points are near to be coplanar. This special disposition is explained by the geometry of the bulky and anisotropic phenyl rings associated to the branching units and the short link between them (in this link, only a O- $CH_2$ bond is not rigidly connected to the ring structures). Our results imply that the ring-O-CH$_2$-ring 4-bond links strongly prefers the "trans" conformation and, consequently, the centers of the four rings adopt a near to coplanar conformation. Therefore, the main contribution to the molecule flexibility is provided by the fluctuations in the orientation of every ring with respect to the bond joining it to the methylene unit of their preceding branch. This orientation automatically defines the orientation of the bonds joining the ring to the oxygen atom in the two successive branches.

When an angle distribution consistent with this description is included as input in the Monte Carlo calculations, the results differ significantly from those calculated without any restriction for the distribution of the angles between branching points. Specifically, the relatively large radius of gyration obtained experimentally for the $g$=3 dendrimer [11] is only reproduced when the angle distribution is introduced in our model. On the other hand, in the case of the monodendrons with higher generation number, our lower choices of $\sigma$ ($\sigma$=1.2 Å or smaller) for the MC simulations with angular distribution always yield values for the radius of gyration that are noticeably above the experimental data (see Table 1). The intrinsic viscosity results are represented by the stars in Figs. 4 and 5. It can be observed that, with this modification, the performance of the $[\eta]$ simulation data is reasonable. However, we should remark that, in this particular case, the best agreement with the experimental data is obtained with two slightly different values of the bead frictional radius, $r_f$=2.37 Å for monodendrons and $r_f$=2.60 Å for tridendrons. Due to all these remaining deficiencies, we have explored other features that may have a substantial contribution in the MC calculations, yielding a more consistent reproduction of all the different experimental data. Nevertheless, we should also point out that, compared with the MC data for the intrinsic viscosity obtained previously without the single bead correction (open circles), a moderately better agreement with the experimental data is actually found for the monodendrons at low $g$ and for the tridendrons at high $g$ when both the single bead correction and the angle distribution are considered.

Investigating the influence of other simple modifications for the PBzE model, we have also verified that the important differences shown in Fig. 1 for the distance distributions obtained for the dendrimer alone and for the system with explicit solvent have a non-negligible influence on the MC final results. In particular, the distance distribution obtained with explicit solvent gives a significantly closer description of the experimental data for the radius of gyration and intrinsic viscosity. In Figs. 4 and 5 we include the results MC results obtained considering this distribution (filled diamonds). A remarkable agreement is observed between simulation and experimental data of monodendrons and tridendrons, both with a common value of the friction radius $r_f$=2.60 Å. In most cases the differences between the two sets of data are within the error bars. Only the simulation point for the $g$=3 monodendron shows a discrepancy with the experimental result, though an improvement with respect to the performance of the original model is also observed for this particular dendrimer. As we have been describing in the preceding paragraphs, the simulation calculations now include the individual bead correction corresponding to this friction radius, the realistic angle distribution and the distance distribution extracted from a dendrimer-solvent system. The hard-spheres distance parameter $\sigma$ in the MC simulations was set to the value $\sigma$=0.5 Å, which adequately also reproduces the existing experimental radius of gyration for the different generation numbers for the monodendrons [11] (see Table 1). This value is very small and contrast strongly with the large value fixed in our previous work, $\sigma$=3.5 Å, that reproduced more closely the radius of gyration when the realistic distribution of angles was not included. We have verified that a moderate increase of the value of $\sigma$ in our present calculations yields a substantial overestimation of the radius of gyration for the $g$=4-5 dendrimers, though the reproduction of the viscosity data is similarly satisfactory.

The near to coplanar disposition of every set of four connected branching points, previously discussed, introduces a significant degree of rigidity in the molecules. It seems that, for our present model based in ideal spherical beads, most of the intramolecular repulsions are effectively mimicked by the rigidity effects and, therefore, the parameter representing these repulsions is drastically reduced. A competition between rigidity and long-range repulsions (or "excluded volume") interactions has been commonly observed in the case of partially rigid chains [20]. It could be argued that an attractive-repulsive potential would give a more realistic description of the size of the dendrimer molecules in different solvents. However, we have introduced such a potential in the present model, verifying that attractions are not able to give a significant reduction of the size of the molecules which is mainly determined by the hard-spheres distance as the single relevant long range interaction parameter.

It should be pointed out that the final agreement achieved between the MC and experimental values of $R_g$ is significantly improved with respect to the results obtained with the simpler original method for the $g$=3 monodendrons (Table 1). A 15% discrepancy was previously observed for this particular case [9]. As previously stated, this particular experimental data can only be reproduced by using a sharp angle distribution function. This confirms that the introduction of realistic details of the distributions of distances and angles obtained with atomistic representation of the molecules give a clearly improved description of the experimental data corresponding to the smallest molecules, poorly represented by our original model.

The agreement between experimental data and the simulation results obtained with all the refinements for the monodendrons and tridendrons of PBzE in THF (filled circles and diamonds, respectively) is significantly better than the accordance found with our previous method (open circles) that did not include any of the refinements considered now, as it can be observed in Fig. 4 and 5. In the particular case of the tridendrons, the experimental data [15] show a sharp maximum in the intrinsic viscosity experimental data and the whole curve is now closely reproduced by the simulation results. As in the case of the monodendrons, the results obtained in our previous work without the present refinements showed a fair agreement with the experimental data, but failed mainly in the prediction of the experimental data corresponding to the smallest and highest values of $g$. It should also be mentioned that the value of the friction radius now obtained for the PBzE molecules, $r_f$=2.6 Å, is considerably smaller than the unrealistic result $r_f$=4.65 Å obtained with the previous simpler method. This new value is also very close to the value set to the PPI-DAB dendrimers, as it has been described above, though this type of agreement has to be considered as fortuitous for dendrimers of different chemical structures in different types of solvents.

In summary, the simple introduction of a correction term taking into account the viscosity of an individual bead offers a good description of the experimental intrinsic viscosity data of the PPI-DAB dendrimers with the multi-step simulation method previously proposed [9]. For the PBzE dendrimers, the more rigid link between branching points requires to introduce further molecular details in the method. Together with the single bead correction, we have considered for these molecules a sharp realistic distribution of angles between connected branching points and also a realistic distribution of distances (including solvent effects). When all these improvements are taken into account, a good general description of the available radius of gyration and intrinsic viscosity data is achieved both for monodendrons and tridendrons of PBzE. It should be remarked that the reproduction of the intrinsic viscosity dependence with the generation number is consistent with a prominent maximum observed in the experimental data of the tridendron molecules.

The present results seem to indicate that the introduction of a single bead correction term is needed to reproduce the experimental intrinsic viscosity data of dendrimers with a coarsed-grained model MC method. Including this correction, a simple model with a rough description of the distribution of distances between branching points is probably enough to give a good reproduction of the experimental data in the case of dendrimers with some flexibility in connecting segments (or spacers) of moderate length as in the case of the PPI-DAB dendrimers investigated in this work. As stated in the Introduction, the higher dispersion of the experimental intrinsic viscosity data does not allow us to contrast the benefits of this modification for the PAMAM-DEA molecules. These dendrimers with longer spacers have been the object of many experimental and theoretical studies. The most recent MD studies of PAMAM-DEA dendrimers with a detailed coarsed-grained model [21] including charges have been mainly focused in the determination of the radius of gyration, density profiles and other conformational and dynamic properties and their interaction with bilayers [22].

However, the consideration of an angle distribution between branching points is needed when the connexion between these points have a mainly rigid structure, as in the case of the monodendron and tridendron PBzE molecules. The introduction of distance distributions obtained from MD simulations with explicit solvent leads to closer MC results for these molecules and this may be also true for other dendrimers including planar rings or other anisotropic units. Of course, these general conclusions require further verification using other dendrimers of similar characteristics, for which accurate experimental data of the intrinsic viscosity data may be available.

**Acknowledgement**. This work has been supported by Projects CTQ2006-06446 and CTQ2006-06831 (including FEDER funds) from DGI-MEC Spain, and the Marie Curie Action MERG-CT-2004-006316. ER and GRE acknowledge predoctoral fellowships associated to the previous DGI-MEC Projects BQU2002-04626C02 and BQU2003-04517, respectively. JGHC acknowledges the award of a "Ramón y Cajal" postdoctoral contract.

Table 1. Mean radius of gyration (in Å) for PBzE molecules in THF.

| $g$ | Monodendrons | | | | Tridendrons | |
| --- | --- | --- | --- | --- | --- | --- |
| | MC simulation | | | | MC simulation | |
| | a) | b) | c) | Experimental | a) | b) |
| 2 | 5.9 | 7.1 | 7.1 | | 7.5 | 8.0 |
| 3 | 8.5 | 9.7 | 9.8 | 9.9 | 10.0 | 10.9 |
| 4 | 11.1 | 12.3 | 12.0 | 11.6 | 12.7 | 13.4 |
| 5 | 13.9 | 14.8 | 13.8 | 13.5 | 15.9 | 15.9 |
| 6 | 17.1 | 16.5 | 17.3 | | 19.1 | 18.6 |
| 7 | 20.8 | 20.6 | 19.2 | | 23.0 | 21.1 |

a) From [9]

b) MC results including distribution of angles, $\sigma$=1.2 Å.

c) MC results including distribution of angles and distribution of

distances obtained from MD simulations with explicit THF solvent,

$\sigma$=0.5 Å.

Statistical error bars are always smaller than 0.2 Å.

**Figure Captions**

Fig. 1.- Distribution of distances between the non-central branching points and ends corresponding to a $g$=2 PBzE dendrimer, expressed as probability per Å, $p(b_c)$. These results have been obtained from MD simulations for the following systems: a) single dendrimer, solid lines; b) dendrimer in TFH, dash line.

Fig. 2.- Distribution of branching point angles between six interval of cos($\Theta$) used as input for the MC simulations, expressed as the normalized probability in each one of these intervals, $p$ (smooth line). The distribution has been estimated from MD results for the dendrimer with g=2, stars; g=3, x; $g$=4, +.

Fig. 3.- Simulation and experimental (filled circles) intrinsic viscosity results for the PPI-DAB dendrimers. The simulation data correspond to the MC conformational sample reported in [9], σ=4.0 Å . Open circles: results from [9], $r_f$=4.60 Å; open squares, results from [9], $r_f$=2.55 Å; filled diamonds: results with the single bead correction, $r_f$=2.65 Å. The line is a spline showing the general trend of the experimental data.

Fig. 4.- Simulation and experimental (filled circles) intrinsic viscosity results for the PBzE monodendrons. The simulation data correspond to different types of calculations. Open circles: results from [9], σ=4.0 Å, $r_f$=4.65 Å; open squares: results from the conformational sample of [9], but with the single bead correction, $r_f$=2.65 Å; stars: results with the single bead correction and the distribution of angles between branching points, σ=1.2 Å, $r_f$=2.37 Å; filled diamonds: results with the single bead correction, the distribution of angles between branching points and a distribution of distances calculated from explicit solvent MD, σ=0.5 Å, $r_f$=2.60 Å. The solid and dash lines are splines corresponding to the experimental data and the MC with only the single bead correction, respectively, and show the general trend of these data.

Fig. 5.- Simulation and experimental (filled circles) intrinsic viscosity results for the PBzE tridendrons. The simulation data correspond to different types of calculations. Open circles: results from [9], , σ=4.0 Å, $r_f$=4.65 Å open squares: results with only the single bead correction, $r_f$=2.65 Å; stars, results with the single bead correction and the distribution of angles between branching points, σ=1.2 Å, $r_f$=2.60 Å; filled diamonds: results with the single bead correction, the distribution of angles between branching points and a distribution of distances calculated from explicit solvent MD, σ=0.5 Å, $r_f$=2.60 Å. The solid and dash lines are splines corresponding to the experimental data and the MC with only the single bead correction, respectively, and show the general trend of these data.

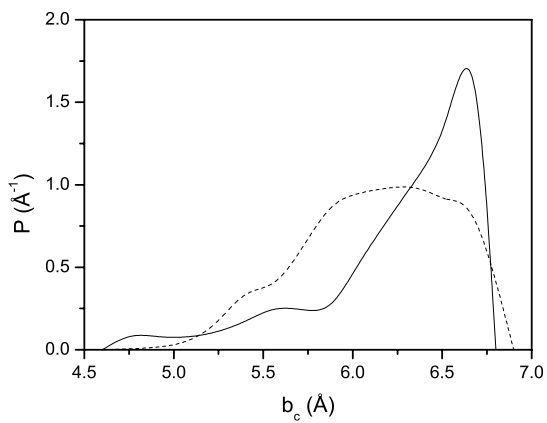

Fig. 1

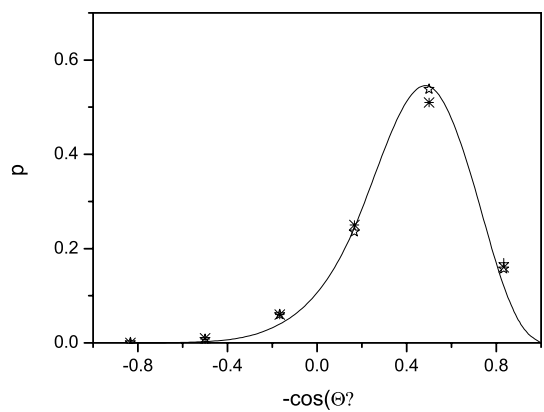

Fig. 2

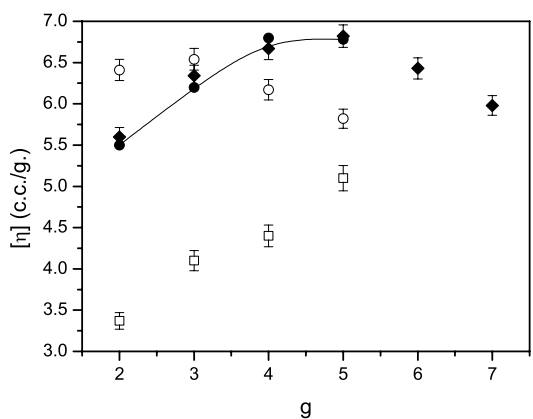

Fig.3

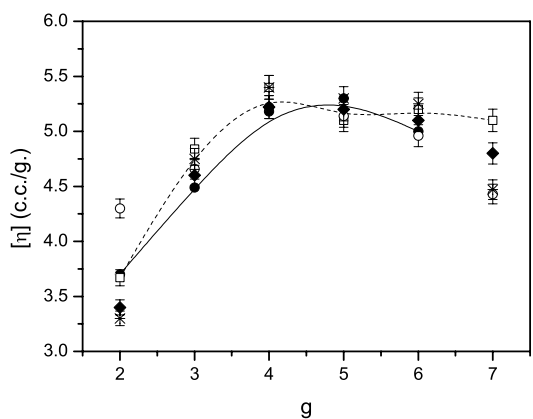

Fig. 4

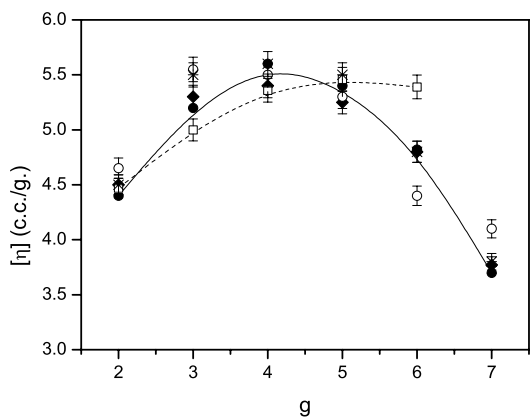

Fig.5